# A network model for clonal differentiation and immune memory

Alexandre de Castro[a,b,*]

[a]*Centro Regional VI, Universidade Estadual do Rio Grande do Sul 90010-030, Porto Alegre, RS, Brazil*
[b]*Instituto de Física, Universidade Federal do Rio Grande do Sul 91501-970, Porto Alegre, RS, Brazil*

**Abstract**

A model of bit-strings, that uses the technique of multi-spin coding, was previously used to study the time evolution of B-cell clone repertoire, in a paper by Lagreca, Almeida and Santos. In this work we extend that simplified model to include independently the role of the populations of antibodies, in the control of the immune response, producing mechanisms of differentiation and regulation in a more complete way. Although the antibodies have the same molecular shape of the B-cells receptors (BCR), they should present a different time evolution and thus should be treated separately. We have also studied a possible model for the network immune memory, suggesting a random memory regeneration, which is self-perpetuating.

*Keywords:* Immune memory; Antibodies evolution; Memory regenerating

## 1. Introduction

In the last decades, due in part to the spread of the world AIDS epidemy, the scientific community started to dedicate special attention to the immunological system, responsible for the defense of the organism against microscopic invaders.

*Corresponding author. Centro Regional VI, Universidade Estadual do Rio Grande do Sul 90010-030, Porto Alegre, RS, Brazil.

*E-mail address:* alexandre-castro@uergs.edu.br.

The immune system is comparable to a sophisticated defense army: the soldiers are highly specialized cells that attack the invaders with potent biochemical tools. The understanding of the defense system of the mammals, specially the immune human system, is very important to find the cure for diseases and to improve the quality of life of the human being.

To control each type of infection, it is necessary a great variety of immune responses. The immune response is mediated mainly by the B and T lymphocytes, responsible for the specific recognition of the antigen [1–3]. In general the immune system presents virgin, immune and tolerant states, and can also present memorization limits. In the virgin state, the amount of B-cells are of the order of those produced by the bone marrow. In the immune state, the population of B-cells that specifically recognizes a type of antigen stays in a certain level after the suppression of the antigen. In the tolerant state, the population of antibodies or B-cells does not respond to antigens or to self-antigens present in our body.

In the human body, there exist millions of B-lymphocytes types [4–7], each one imprisoning, in its membrane, the specific antibody. From all these lymphocytes, just those that recognize an antigen are stimulated. In this case, the stimulated B lymphocytes, originates a lineage of cells (by multiplication or differentiation process) which is capable of producing specific antibodies against the antigen that induced the multiplication. This phenomenon is called clonal selection.

The antibodies produced by a mature B lymphocyte, are liberated in large quantities in the blood. The multiplication continues as long as there are antigens capable of activating them. When a certain type of antigen is eliminated from the body, the number of lymphocytes specialized in combat also decreases. However, a small population of lymphocytes remains in the organism for the rest of the life of the individual, constituting what is denominated *immune memory*. In the case of the same antigen invading again the organism, this remaining population multiplies quickly. Then, in a small interval of time, a great amount of B lymphocytes produces antibodies, in order to quickly combat the invader. This explains why the production of antibodies in the secondary immune response is faster than in the primary response.

There are basically two theories to explain the immune memory. The first considers that, after the expansion of the B-cells, there occurs the formation of plasma and memory cells [8]. These memory cells are the cells remaining from an immune response, that survive until the end of the individual's life—therefore with longer life than the other cells of the organism.

The second theory, due to Jerne [9,10], considers that the immune system presents its memory and its response capacity to the second invasion of antigens as a self-organization of the system, allowing the formation of cell populations that survive for a long time, but not longer than the other cells of the organism.

Different models based on the clone dynamics have been proposed to describe the interaction between different cells, their proliferation and death, antibody secretion and interaction with antigens. The Celada–Seiden model [11–14], for example, is one of the most detailed automata for the immune system response. Its complexity derives from the fact that, in addition of considering the different cellular populations, a representation of the molecular binding site is also given in terms of specific recognition between bit strings. The match between antigens and lymphocyte receptors is given, considering the number of complementary bits in the bit-strings. For example, if the B lymphocyte is equipped with the binary string 00010101 ($\ell = 8$) and the antigen is represented by the string 11101010, then the probability of triggering a response is very high. In this model the bit-string match is not required to be perfect, some mismatches are allowed.

In this paper, we have developed a bit-string model for the immune system, considering structural mechanisms of regulation that were not contemplated in the simplified model of Lagreca et al. [15]. In our approach we take into account not only the antibodies bound to the surface of the B-cells (surface receptors), but also the populations of antibodies soluble in the blood (antibodies secreted by mature B-cells), making this model closer to a real immune system. The concentrations (doses) of the entities are treated following the biological nomenclature.

In order to study more appropriately the time evolution of the components of the immune system, we define clone as an ensemble of B-cells only. Therefore, the antibody population are treated separately in this work.

On the other hand, the results obtained by Lagreca et al. [15] for the time evolution of the clones only present the time evolution of B-cells with its surface receptors and not the evolution of the populations of secreted antibodies. This is due to the fact that, in their model, the set of coupled maps considers the antibodies bound to the surface of B-cells and not the antibodies dispersed in the serum.

Moreover, the regulation of the immune response in the previous model is just by apoptosis (programmed cell death) and by the Verhulst-like factor. It has not being considered the fundamental role of the antibodies in the mediation of the global control of the differentiation of the B-cells.

Our model allows to represent the generation, maintenance and regulation of the immune memory in a more complete way, through a network memory, combining the characteristics of Burnet's clonal selection theory and Jerne's network hypothesis, considering only the idiotypic–antiidiotypic interaction. The combination of these two theories lead to the hypothesis of relay of Nayak et al. [16], that does not require the presence of long-living memory cells or persisting antigens. We have obtained results for numerical simulations of the Nayak model, considering also a modified version of exact enumeration techniques, together with multispin coding [17–20], to allow the control of time evolution of populations over high-dimensional shape space.

This paper is organized as follows: in Section 2 we describe the theoretical model and its entries; in Section 3 we present results of our computer simulations for the time evolution of populations in the immune system during the infections; and in Section 4 we discuss our improved model, that includes antibodies populations into the system of coupled maps.

## 2. The computational model and biological entities

In the model discussed here, the B-cell molecular receptors are represented by bit-strings. Each clonotypic set of cells has a diversity of $2^B$, where $B$ is the number of bits in the strings.

The individual entities present in the model are B-cells, antibodies and antigens. The B-cells are represented by clones, that are characterized by its surface receptor, modeled by a binary string.

The epitopes [21,22] are also represented by bit-strings. Epitope is the portion of an antigen that can be bound by the B-cell receptor (BCR). The antibodies have a receptor (paratope) [4] that is represented by the same string as the BCR of the parent B-cell which produced them.

Each bit-string (shape) is associated to an integer $\sigma$ ($0 \leqslant \sigma \leqslant M = 2^B - 1$) that represent each clone, antigen or antibody. The neighbors to a given $\sigma$ are expressed by the Boolean function $\sigma_i = (2^i x \text{ or } \sigma)$. The complementary shape of $\sigma$ is obtained as $\overline{\sigma} = M - \sigma$ and through direct iteraction, the time evolution of the concentrations of several populations is obtained as a function of integer variables $\sigma$ and time $t$.

The equations that describe the behavior of the populations of B-cell clones $y(\sigma, t)$ are iterates, for different parameters and initial conditions:

$$y(\sigma, t+1) = (1 - y(\sigma, t))\left\{ m + (1-d)y(\sigma, t) + b\frac{y(\sigma, t)}{y_{TOT}(t)}\zeta_{a_h}(\overline{\boldsymbol{\sigma}}, t) \right\} \tag{1}$$

with the complementary shapes included in the term $\zeta_{a_h}(\overline{\boldsymbol{\sigma}}, t)$

$$\begin{aligned} \zeta_{a_h}(\overline{\boldsymbol{\sigma}}, t) = {} & (1 - a_h)(y(\overline{\sigma}, t) + y_F(\overline{\sigma}, t) + y_A(\overline{\sigma}, t)) \\ & + a_h \sum_{i=1}^{B} (y(\overline{\sigma}_i, t) + y_F(\overline{\sigma}_i, t) + y_A(\overline{\sigma}_i, t)) \,, \end{aligned}$$

where $y_A(\sigma, t)$ and $y_F(\sigma, t)$ are the antibody and antigen population, respectively. Here $b$ is the proliferation rate of B-cells, $\overline{\sigma}$ and $\overline{\sigma}_i$ are the complementary shape to $\sigma$ and to the $B$ nearest-neighbors on the hypercube (with the $i$th bit flipped). The first term in the curled bracket ($m$) represents the bone marrow production, and is a stochastic variable. This term is small, but non-zero. The second term describes the population that survives to the natural death ($d$), and the third term represents the clonal proliferation due to the interaction with complementary forms (other clones, antigens or antibodies). The parameter $a_h$ is the relative connectivity among a given bit-string and the neighborhood of its mirror image. When $a_h = 0$, only perfect match is allowed. When $a_h = 0.5$, a string may recognize equally well its mirror image and first neighbors.

The factor $y_{TOT}(t)$ is given by

$$y_{TOT}(t) = \sum_{\sigma} [y(\sigma, t) + y_F(\sigma, t) + y_A(\sigma, t)] \,. \tag{2}$$

The time evolution of the antigens is given by

$$\begin{aligned} y_F(\sigma, t+1) = {} & y_F(\sigma, t) - k\frac{y_F(\sigma, t)}{y_{TOT}(t)} \\ & \times \left\{ (1 - a_h)[y(\overline{\sigma}, t) + y_A(\overline{\sigma}, t)] + a_h \sum_{i=1}^{B} [y(\overline{\sigma}_i, t) + y_A(\overline{\sigma}_i, t)] \right\} , \end{aligned} \tag{3}$$

where $k$ is the speed with which populations of antigens or antibodies decay to zero. Here it measures the removal rate of the antigen due to interaction with the population of clones and antibodies.

The population of antibodies is described by a group of $2^B$ variables, defined on the B-dimensional hypercube, interacting with the populations of antigens:

$$y_A(\sigma, t+1) = y_A(\sigma, t) + b_A \frac{y(\sigma, t)}{y_{TOT}(t)} \times \left[ (1 - a_h) y_F(\overline{\sigma}, t) + a_h \sum_{i=1}^{B} y_F(\overline{\sigma}_i, t) \right] - k \frac{y_A(\sigma, t)}{y_{TOT}(t)} \zeta_{a_h}(\overline{\boldsymbol{\sigma}}, t) , \quad (4)$$

where the complementay shapes contribution $\zeta_{a_h}(\overline{\boldsymbol{\sigma}}, t)$ is again included in the last term. Here, $b_A$ is the proliferation rate of antibodies; $k$ is the antibodies removal rate that measures its interactions with other populations.

The antibodies populations $y_A(\sigma, t)$ (total number of antibodies) depend on the inoculated antigen dose. The factors $y_F(\sigma, t)/y_{TOT}(t)$ and $y_A(\sigma, t)/y_{TOT}(t)$ are the responsible for the control and decay of the antigens and antibodies populations, while $y(\sigma, t)/y_{TOT}(t)$ is the corresponding factor for the accumulation of the clone population in the immune memory building. The clone population $y(\sigma, t)$ (the normalized total number of clones) may vary from the bone marrow value ($m$) to its maximum value (in our model, the unity) since the Verhulst-like factor limits its growth.

The Verhulst-like factor produces a local control of the populations of clones (B-cells) considering the various mechanisms of regulation. However, globally the immunological memory of the system is strongly affected by the populations of soluble antibodies in the blood. This is the reason that lead us to include $\zeta_{a_h}(\overline{\boldsymbol{\sigma}}, t)$ as an extra contribution in the iterate equations.

Eqs. (1)–(4) form a set of equations that describe the main interactions in the immune system between entities that interact through key-lock match, i.e., that specifically recognize each other. This set of equations was solved iteratively, considering different initial conditions.

## 3. The dynamics

To show the strength of the model, we present the results for some simulations where immunization experiments are reproduced, in which the various antigens, with fixed concentration, are injected in the body at each 1000 time steps, to stimulate the immune response. When a new antigen is introduced, its interaction with all other components in the system is obtained accordingly to the random shape generator chosen.

### 3.1. Parameters

The bit-string length $B$ has been set to 12, corresponding to the potential repertoire of 4096 distinct receptors and molecules. We set 110 injections of different antigens in a range of 0–110 000 time steps.

The value for the rate of the natural death of the cells (apoptosis) was taken as $d = 0.99$, and the rate of clone and antibody proliferation were considered as 2 and 100, respectively. The connectivity parameter $a_h$ was chosen as 0.01. The value of both antigens and antibodies removal rates ($k$) have been set to 0.1, so that in each interval of 1000 time steps, the previous populations of antigens and antibodies vanish before the next antigen be present.

We have used the same seed for the random number (integer) generator for each inoculation, so that the different antigens are inoculated in the same order in all simulations. Several simulations were accomplished, with antigen doses from 0.0001 to 1.5.

In this paper, we will discuss the results obtained for two doses in the intermediate region, 0.08 and 0.10, that, although being quite close values, present distinct evolution for the immune memory. The results for doses in the extreme limits, with very peculiar behavior, will be treated in a following paper.

### 3.2. *Numerical results*

The generalized shape space is a hypercube in a 12-dimensional space with non-negative integer coordinates. The corresponding number of discrete points in the shape space is considered to be an adequate immune repertoire. We have performed simulations with different types of antigens, i.e., several infections with many different pathogens. The immune system quickly begins to respond after the beginning of the simulation.

#### 3.2.1. *Evolution of antibodies and antigens*

Numerical simulations using the proposed model allow to obtain results for the time evolution of the populations of B-cells, antibodies and antigens.

Figs. 1(a) and (b) show the results for the behavior of the populations of antibodies for antigen doses of 0.08 and 0.10, respectively. In the present model, with the populations of antibodies being treated separately (antibodies secreted in an immune response), the result of the time evolution is a sequence of peaks, in agreement with what was expected: for a healthy individual the populations of antibodies should decrease when the infection finishes. From these figures we can observe that the overall behavior for both doses (0.08 and 0.10) is approximately the same: a weak floation with time after many injections of antigens. For the larger concentration, there appears a instability for the injection 82 (in 82 000 time step), but the concentration of antibodies is approximately stable in the rest of the studied time interval. This result was expected, because in a dynamic system with different attractors, the response (antibodies) is not necessarily the same in each point of the shape space.

Figs. 2(a) and (b) show in detail the evolution of the populations of antibodies and antigens, for a range of 0–6000 time steps ($dose = 0.08$). The populations of antigens decrease quickly when antibodies are secreted and the antibodies response is proportional to the dose of antigens inoculated according to the behavior of a normal immune system. Similar behavior was observed for every concentration considered, in each case, only the maximum value of the populations varies, being larger for higher antigen doses.

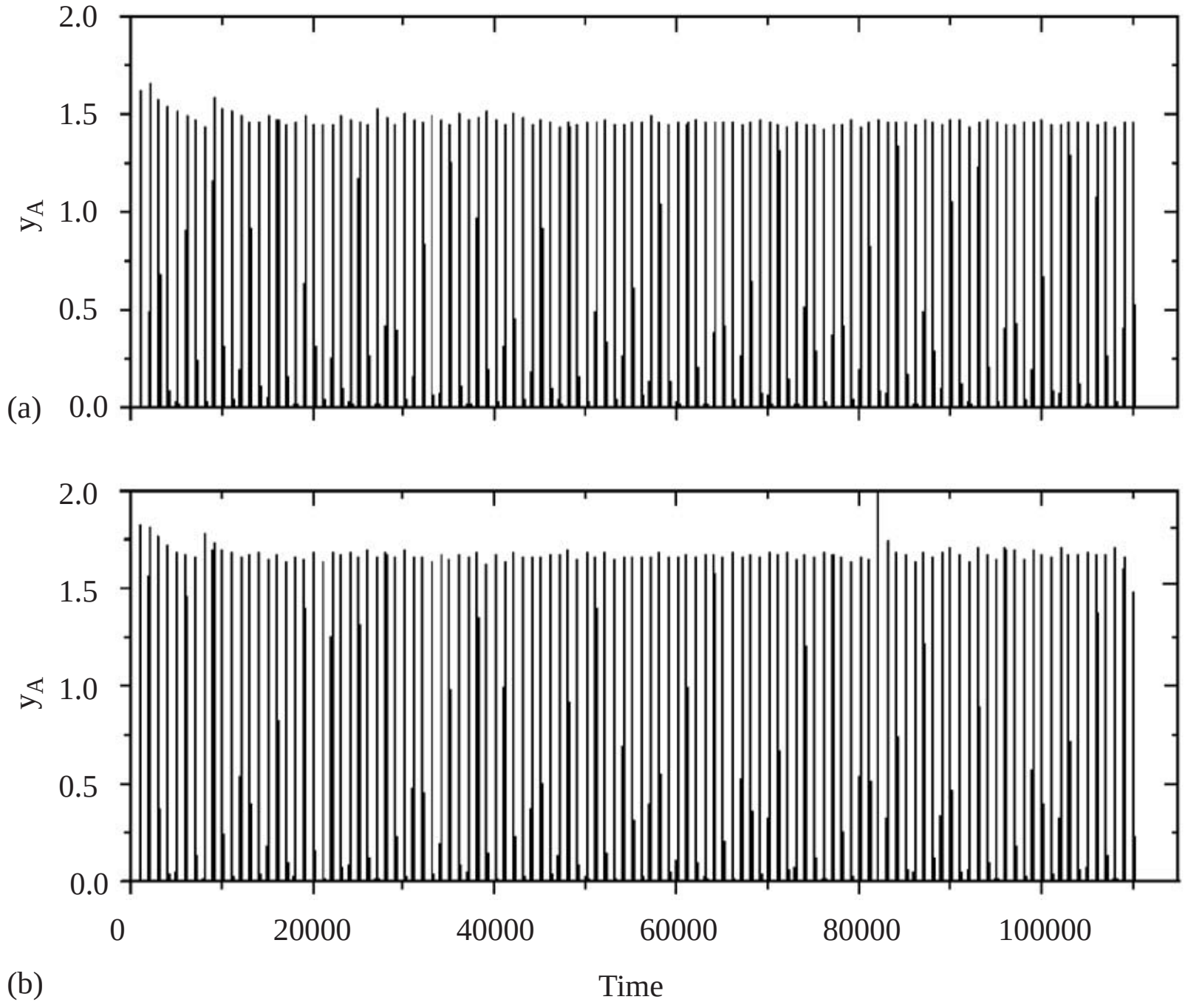

Fig. 1. Model performance of antibodies populations $y_A$ as function of time, for a dose of antigen equal to (a) 0.08 and (b) 0.10. A new antigen population is presented each interval of 1000 time steps (with 110 injections of different antigens populations).

In Figs. 3(a) and 4(a), the time evolution of the population of B-cell clones, averaged over the number of discrete points in the shape space, as shown, for both doses (0.08 and 0.10). Our approach allows to obtain separately the time evolution of B-cells and antibodies. Therefore, the previous result by Lagreca et al. [15] (clone evolution) cannot be understood as the evolution of an ensemble of both B-cells and antibodies, but as being the evolution of B-cells only. The soluble antibodies in the blood should be treated separately because they present behavior different from the evolution of B-cells [1,2]. In a normal immune system the populations of antibodies are not maintained in a fixed level after the ending of the infection. When the populations of antigens decrease, the populations of antibodies also decrease.

### 3.2.2. *Regulation of cell proliferation and memory capacity*

As can be seen in the following, the performed simulations show the generation, maintenance and regulation mechanisms of the immune memory and cellular differentiation, through interactions idiotypic–antiidiotypic combining the characteristics of Burnet's clonal selection theory [8] and immune network theory proposed by Jerne [9,10].

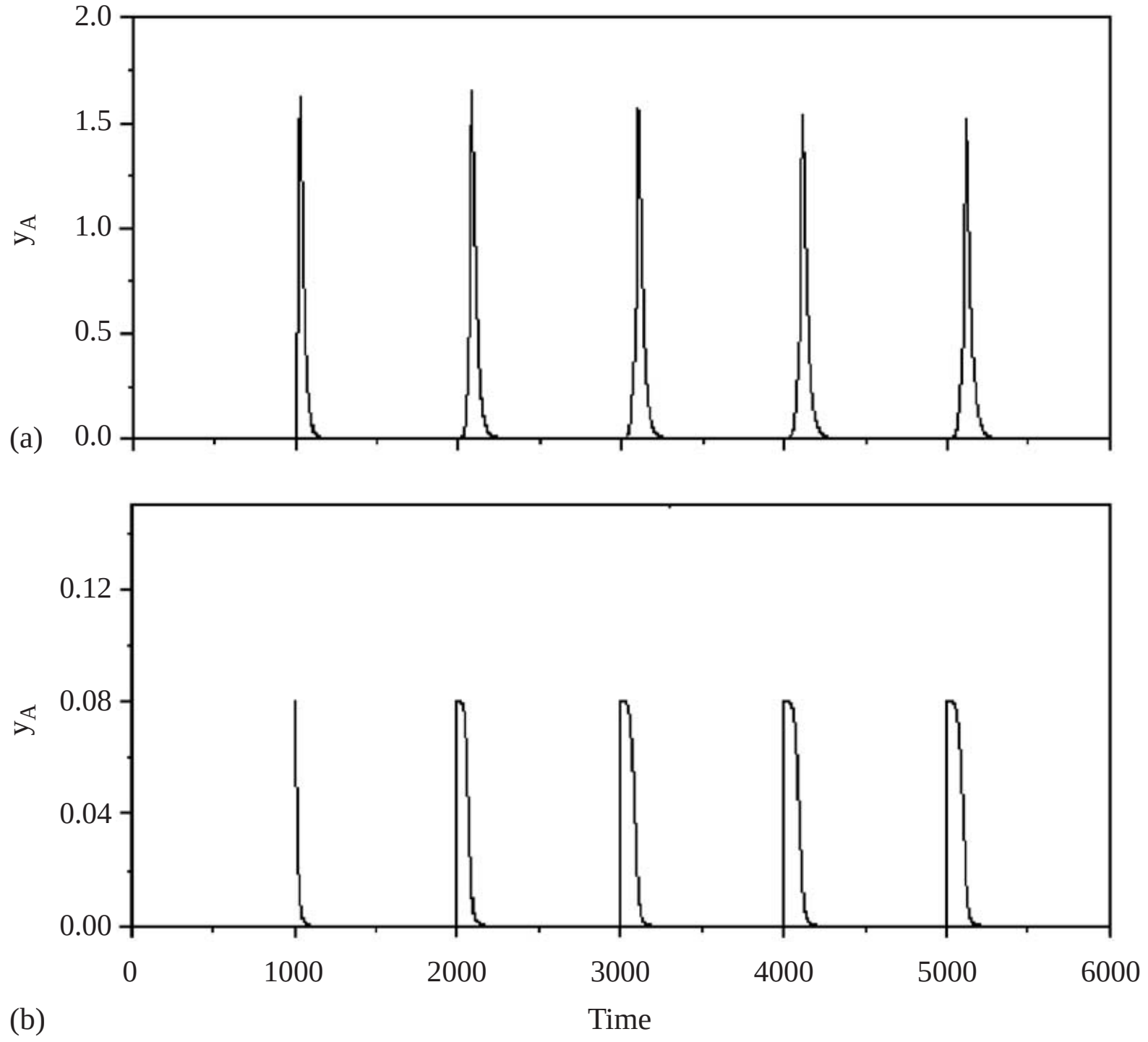

Fig. 2. Model performance of the time evolution (a) of antibodies populations $y_A$ and (b) antigens populations $y_F$ for a range of time steps 0–6000, with concentration of antigens equal to 0.08.

Figs. 3(a) and 4(a) indicate that $\langle y(\sigma)\rangle$, the time evolution of the average concentration of clones (B-cells), increases after some injections of different antigens due to the formation of populations of memory clones. This increase continues until a certain time where the saturation of the evolution is reached, due to the capacity of memorization, because the living body has a maximum number of cells it can support.

Figs. 3(b) and 4(b) show the maximum number of excited populations of clones, after 110 injections, for the two different concentration of antigens. In Fig. 5, the excited clones populations are represented (dose equal to 0.08), for both the first antigen inoculation, Fig. 5(a), and the second, Fig. 5(b). In this evolution, two populations are excited when the first antigen population is inoculated: the clonal population that recognized a specific antigen (Burnet cells) and the clonal population with complementary shapes (Jerne cells).

When the second antigen population is inoculated, we can observe four excited peaks—the complementary shape contribution is also present. Similar result were obtained for dose equal to 0.10.

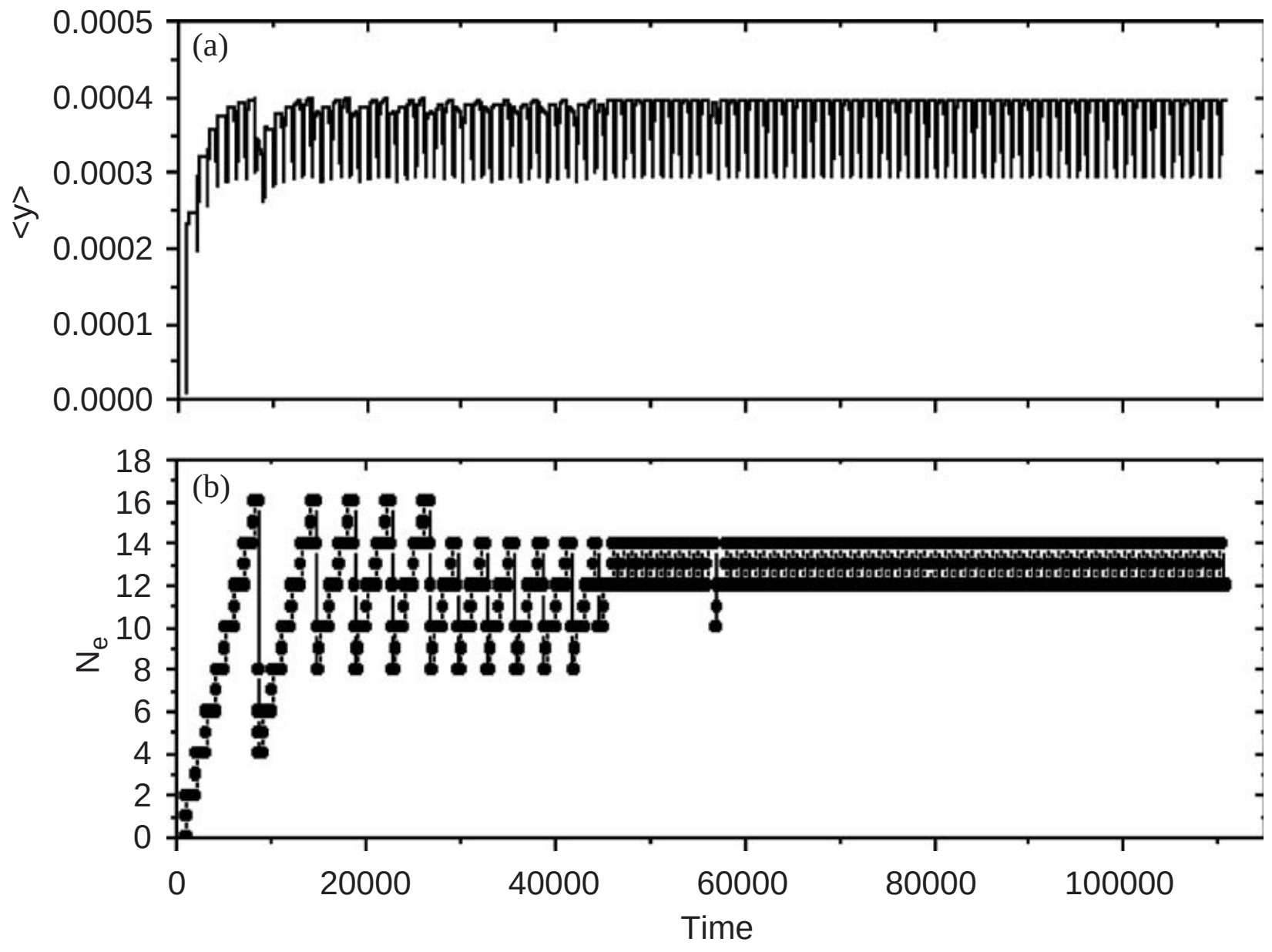

Fig. 3. Model performance for the time evolution for concentration of antigens equal to 0.08. Average clonal population $\langle y(\sigma)\rangle$ (a) and number of excited populations $N_e$ (b).

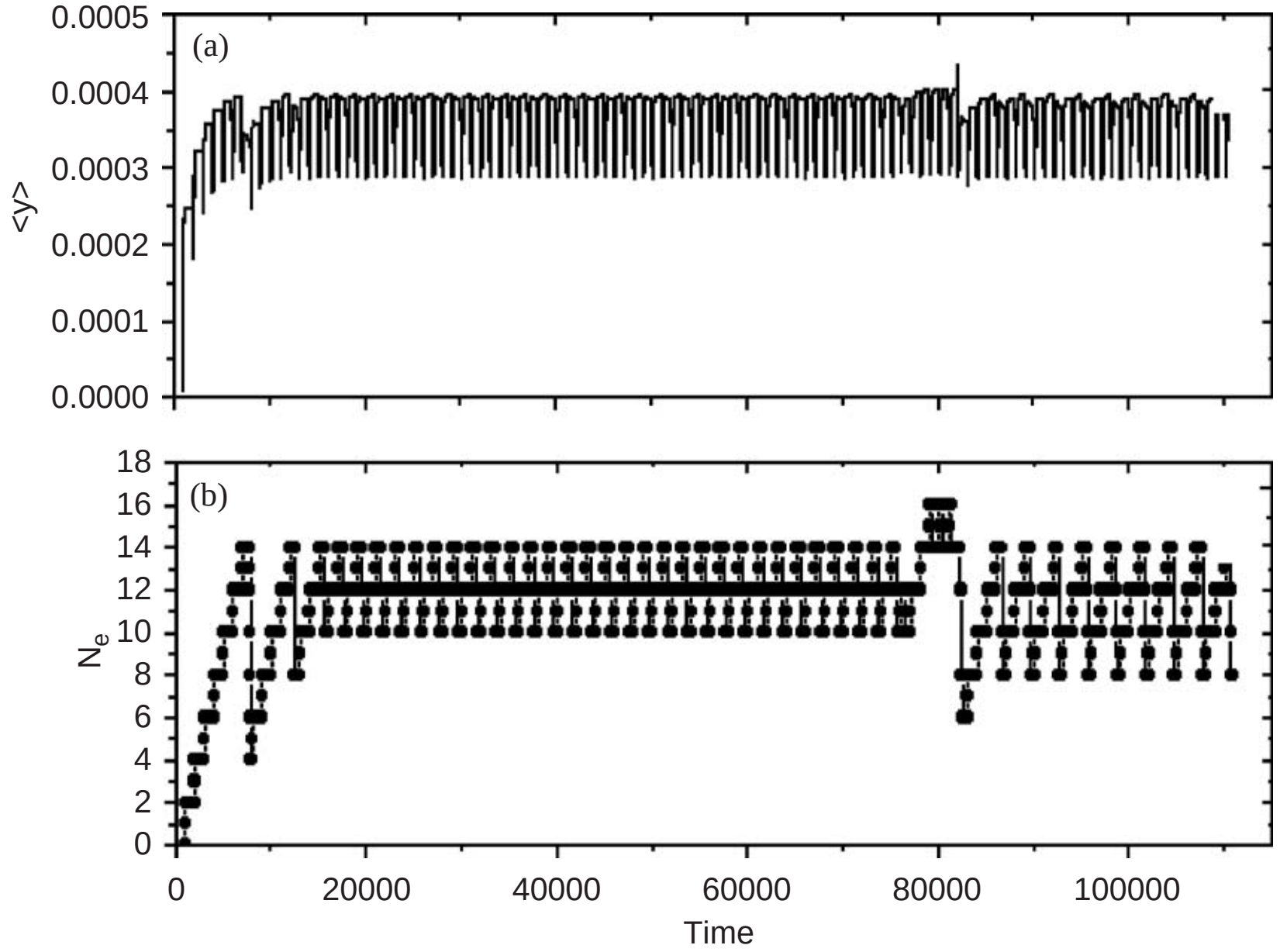

Fig. 4. Model performance for the time evolution for concentration of antigens equal to 0.10. Average clonal population $\langle y(\sigma)\rangle$ (a) and number of excited populations $N_e$ (b).

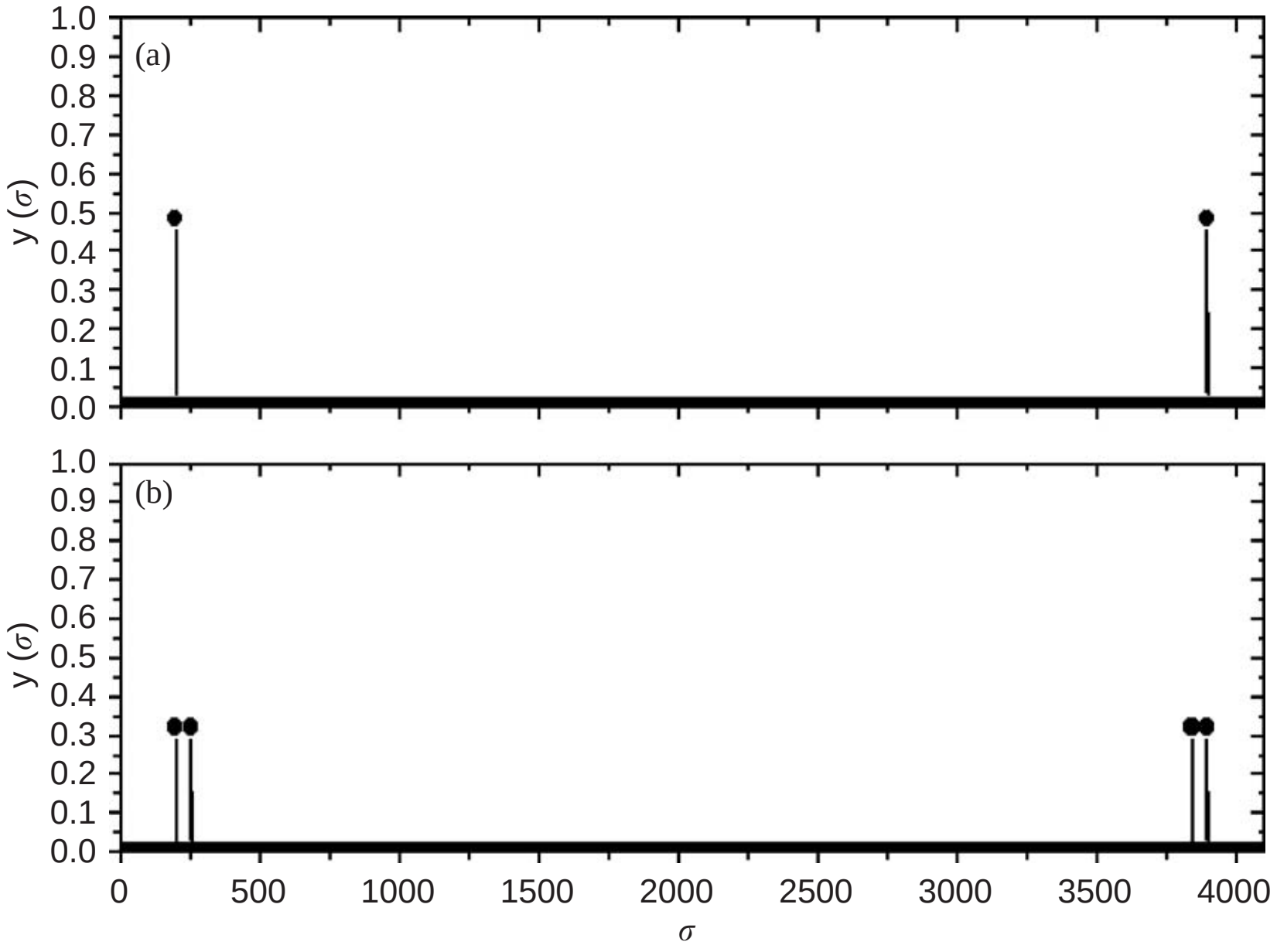

Fig. 5. Excited clones populations (*dose* = 0.08) for the first antigen inoculation (a). The corresponding result for the second inoculation is presented in (b).

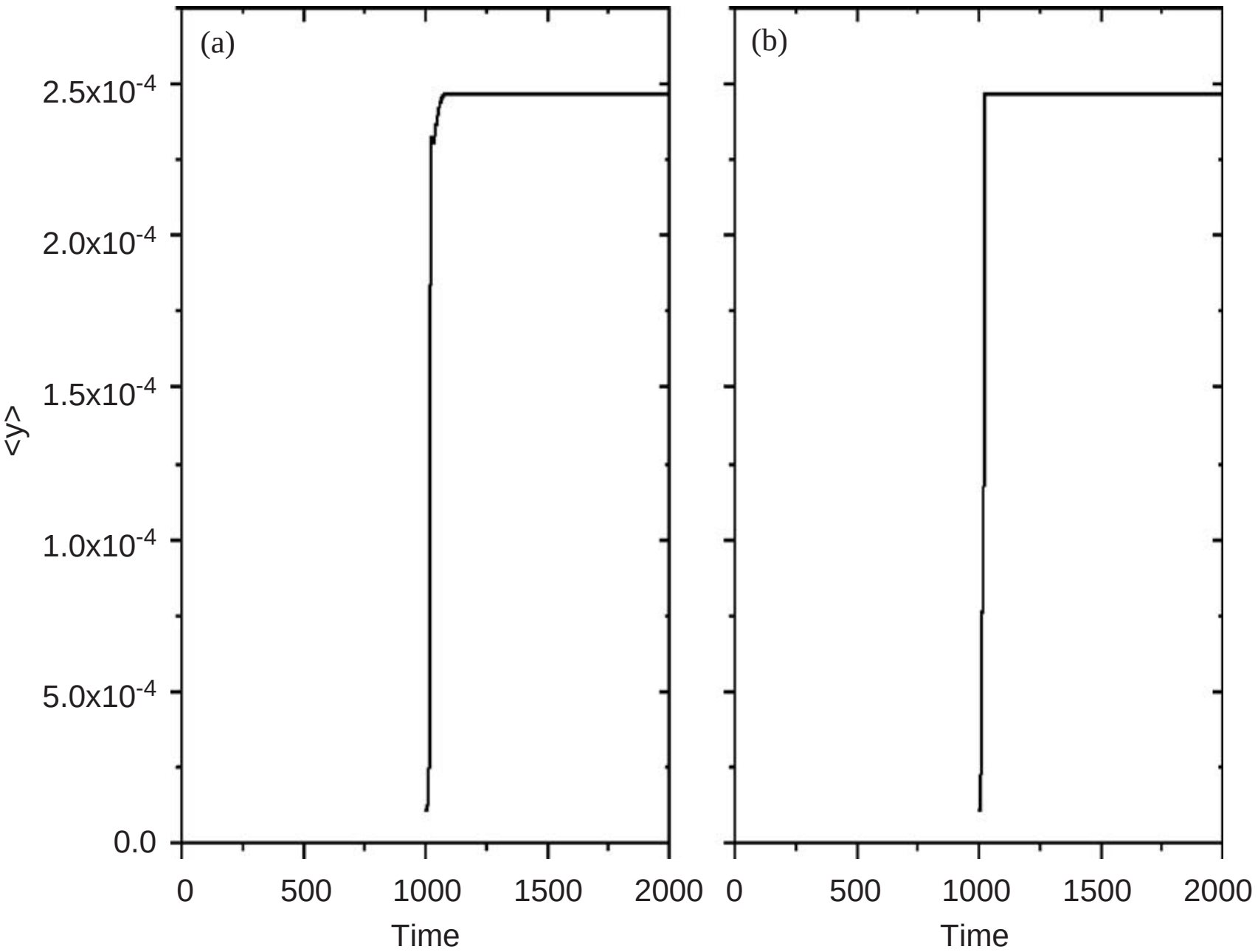

Fig. 6. Time evolution of the first population of clones that recognizes the first inoculated antigen with (a) addition of populations of antibodies and (b) without populations of antibodies. Identical behavior is found for antigens concentrations 0.08 or 0.1.

Figs. 6(a) and (b) show the time evolution of the first population of clones that recognizes the first inoculated antigen: (a) with the addition of populations of antibodies, and (b) without antibodies in the set of coupled maps. These behaviors do not present noteable difference for different antigen concentration. The results obtained without antibodies, Fig. 6(b), correspond to the simplified model of Lagreca et al. [15].

The addition of antibodies in the system does not originate a considerable local perturbation, however, in Figs. 7 and 8 it is shown that the addition of antibodies alters the global memory capacity. In these figures, the memory capacity is represented, considering the system with or without the presence of antibodies. For both antigen concentrations of 0.08 and 0.10, when the antibodies populations are considered, Figs. 7(a) and 8(a), the network capacities are smaller than in the absence of antibodies populations, Figs. 7(b) and 8(b). In the absence of a specific model for the antibodies, the populations reach higher levels. This indicates the important role of the antibodies in the mechanism of regulation of the proliferation of B-cells and in the maintenance of the immune memory. The decrease in the amount of active populations is due to the interaction among antibody and B-cells, in agreement with the immune network theory.

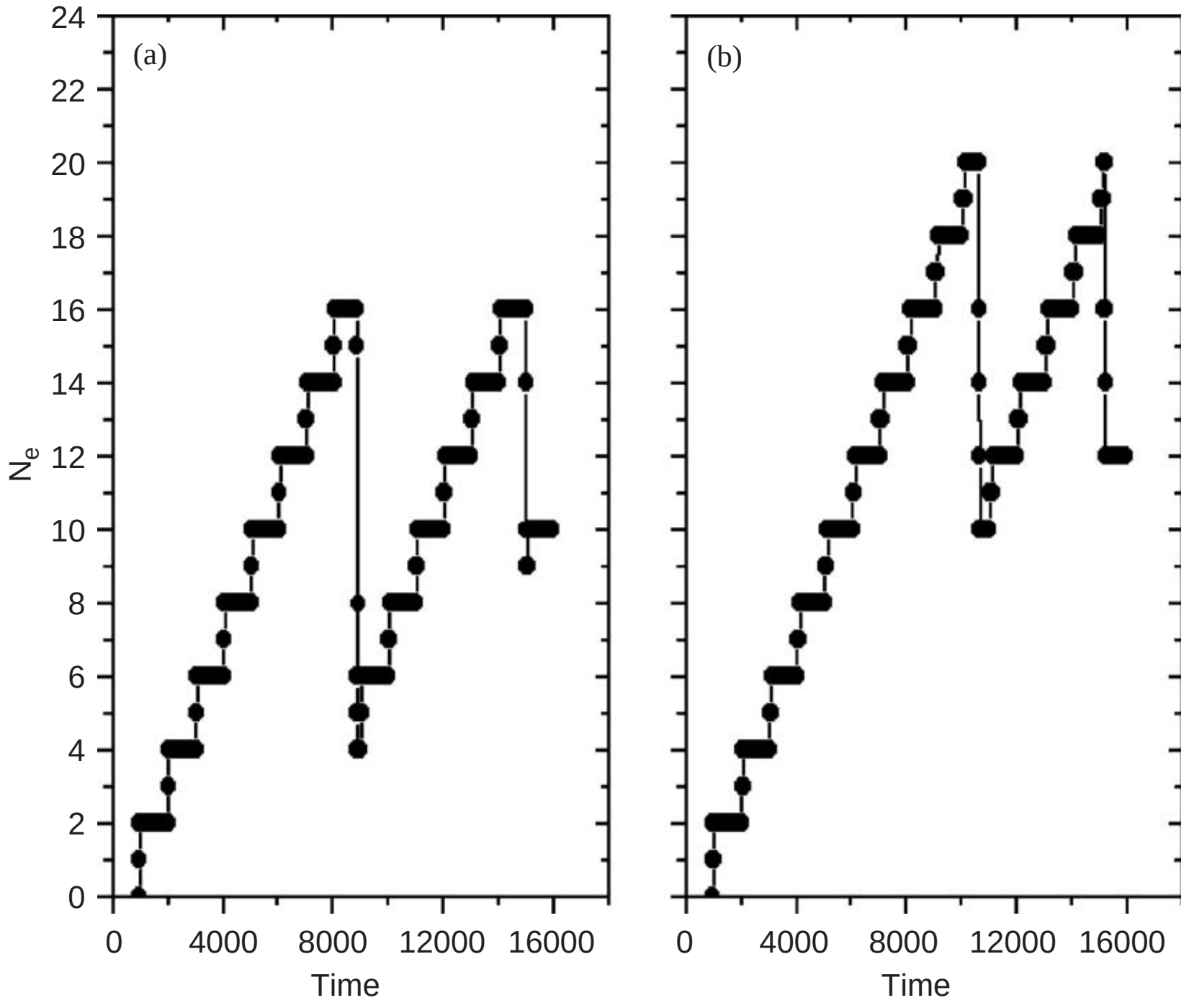

Fig. 7. Memory capacity as a function of time, for the concentration of antigens equal to 0.08. The network capacity with the addition of populations of antibodies (a) is smaller than in (b), without antibodies.

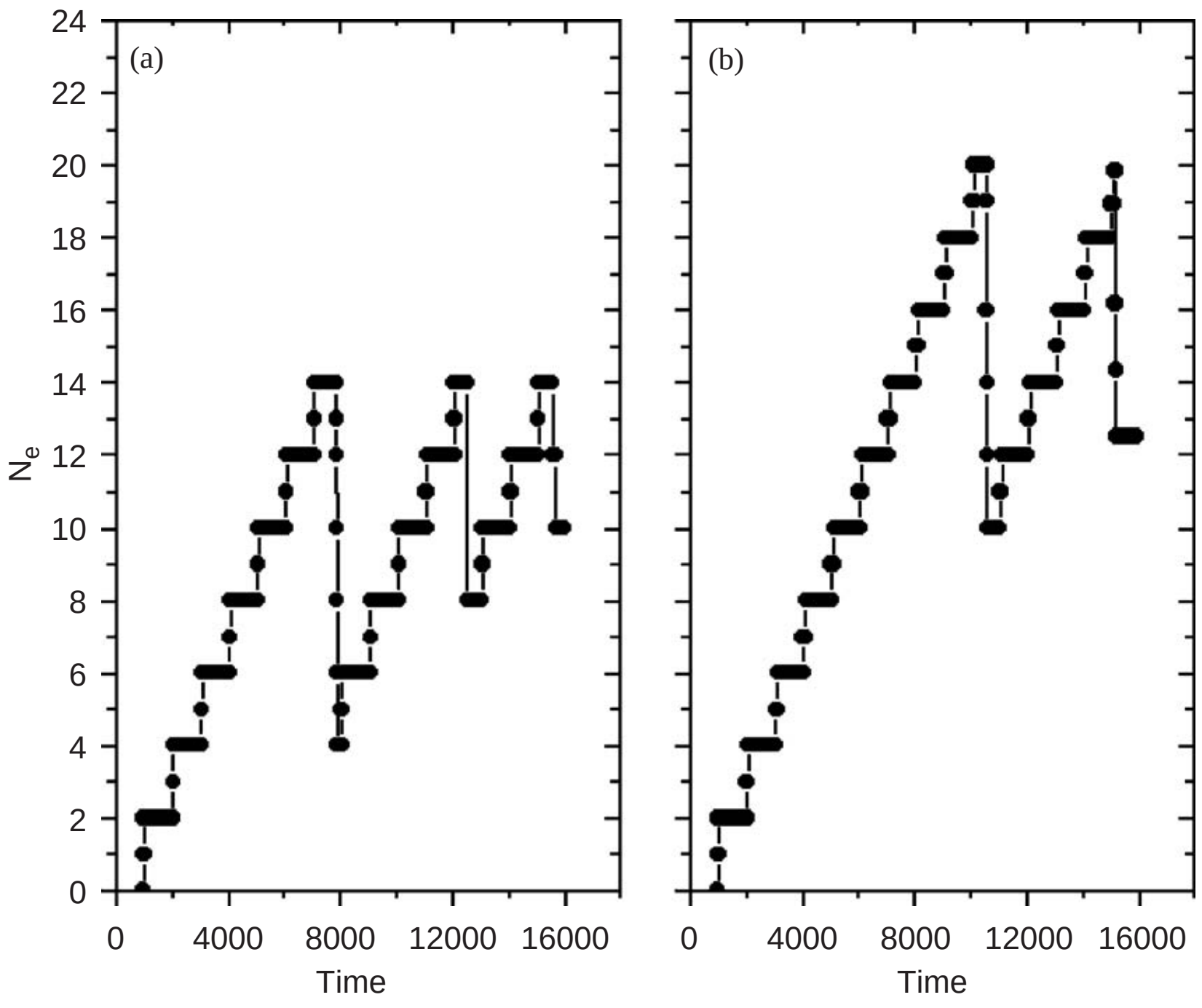

Fig. 8. Memory capacity as a function of time, for the concentration of antigens equal to 0.10. The network capacity with the addition of populations of antibodies (a) is smaller than in (b), without antibodies.

The populations of soluble antibodies in the blood participate not only in the immune response but also help in the regulation of the differentiation of B-cells. The presence of soluble antibodies alters the global properties of the network: this behavior can only be observed when the populations are treated separately. The approach of Lagreca et al. [15] only considers the antibodies bound to the surface of B-cells: it does not take into account those secreted by mature B-cells. The proposed improved model allows to simulate the behavior of the populations of antibodies and to visualize its influence in the mechanisms of regulation of the response and in the memory capacity.

Figs. 9 and 10 show the populations of clones that specifically recognize the antigens, with concentration equal to 0.08. In this case, for a typical simulation until the tenth injection, the fourth 9(d) and eighth 10(c) populations of clones remain active until 11 000 time steps. All the other 8 populations have shorter lives. For the concentration of antigens equal to 0.10, Figs. 11 and 12, the populations surviving for a long period are the first 11(a) and the seventh 12(b), the other 8 remaining populations have shorter life. This behavior is random, being therefore impossible to foresee, for a given antigen dose, which clone population stays excited for a longer time.

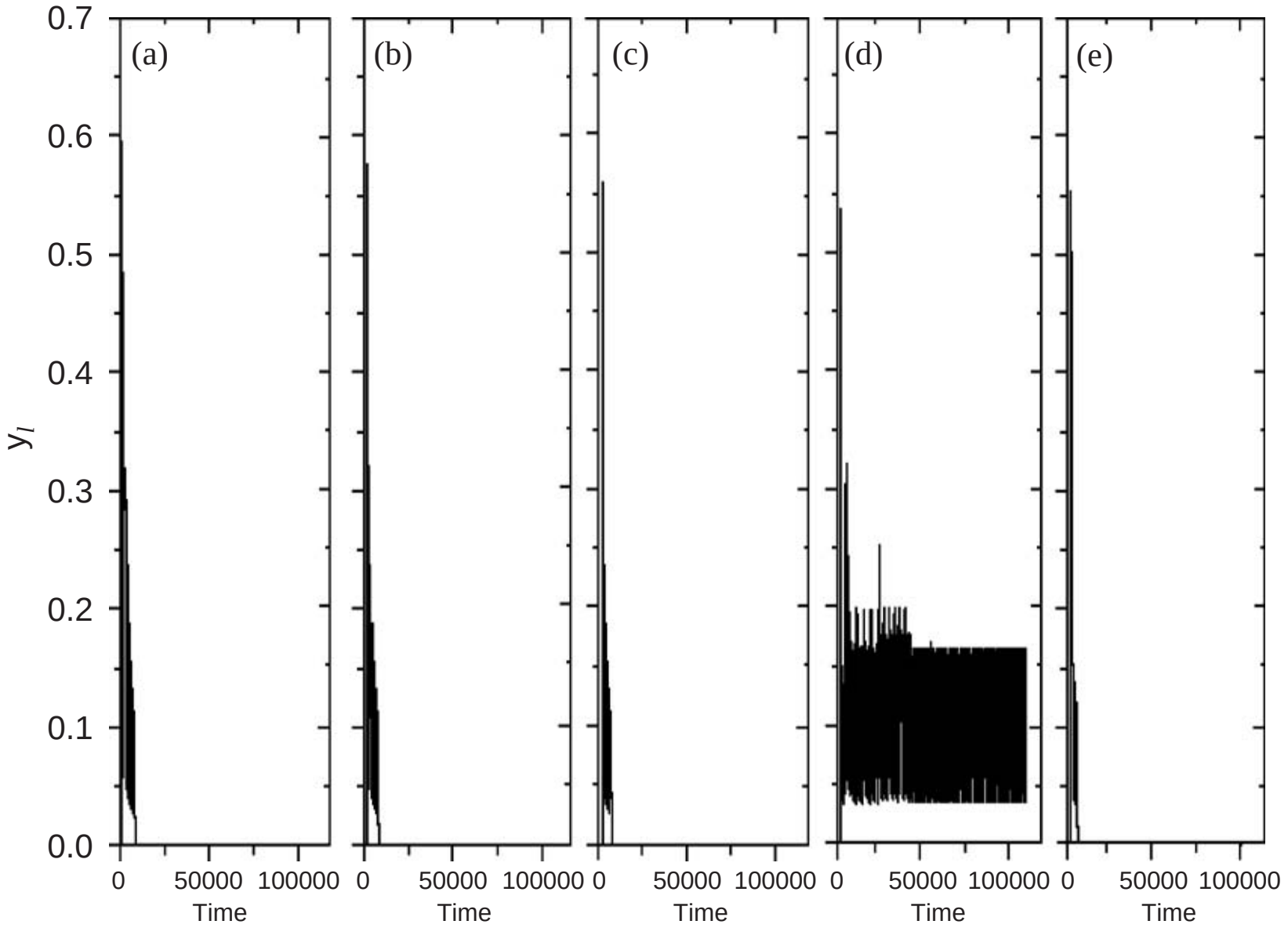

Fig. 9. Performance of the model for the first five populations of clones that recognize a specific population of antigen (concentration equal to 0.08), shown individually (a–e). All the populations (1–5) are excited after the inoculation, however just the fourth population of clones (d) remains for a long time after the suppression of the antigen.

Figs. 13 and 14 show with more detail the evolution of the specific populations, for each time step. These pictures show that while a population is forgotten, another one is learned. We can observe in Fig. 14 that the population that recognized the first type of antigen proliferates to a certain point, beginning to decrease in the second injection. This behavior is due to the fact that the immune system has a maximum capacity of cells it can tolerate. When new types of antigens are learned, others need to be forgotten. However, in the 7000 time step, when the seventh injection is applied, the first population begins to increase. This indicates that the first population could represent a long time memory.

The increase in the concentration of the first population after the 7000 time step shows that our model describes the behavior of the immune network proposed by Jerne [9,10], where the immune memory is formed by populations of cells that last for a long time through cell–cell interactions, and not by a specific type of cell that has a life longer than the other cells of the body.

The proposed model simulates the hypothesis of relay of Nayak et al. [16] combining characteristics of the Burnet's clonal selection theory and Jerne's network hypothesis. In this approach, the presence of long-living memory cells or persisting antigen is not required, however the individual behavior of each population is completely random.

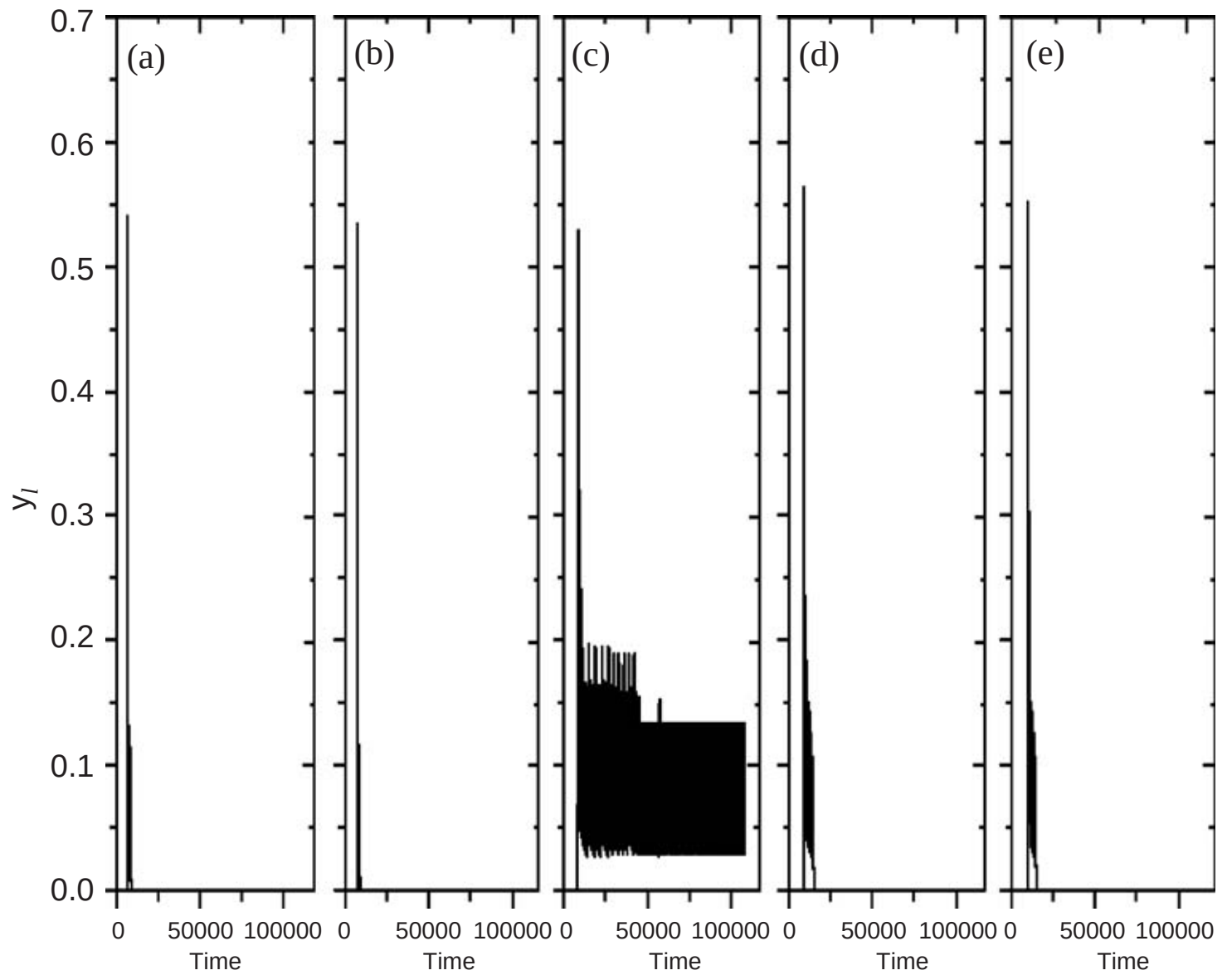

Fig. 10. Performance of the model for the first five populations of clones that recognize a specific population of antigen (concentration equal to 0.08), shown individually (a–e). All the populations (6–10) are excited after the inoculation, however just the eighth population of clones (c) remains for a long time after the suppression of the antigen.

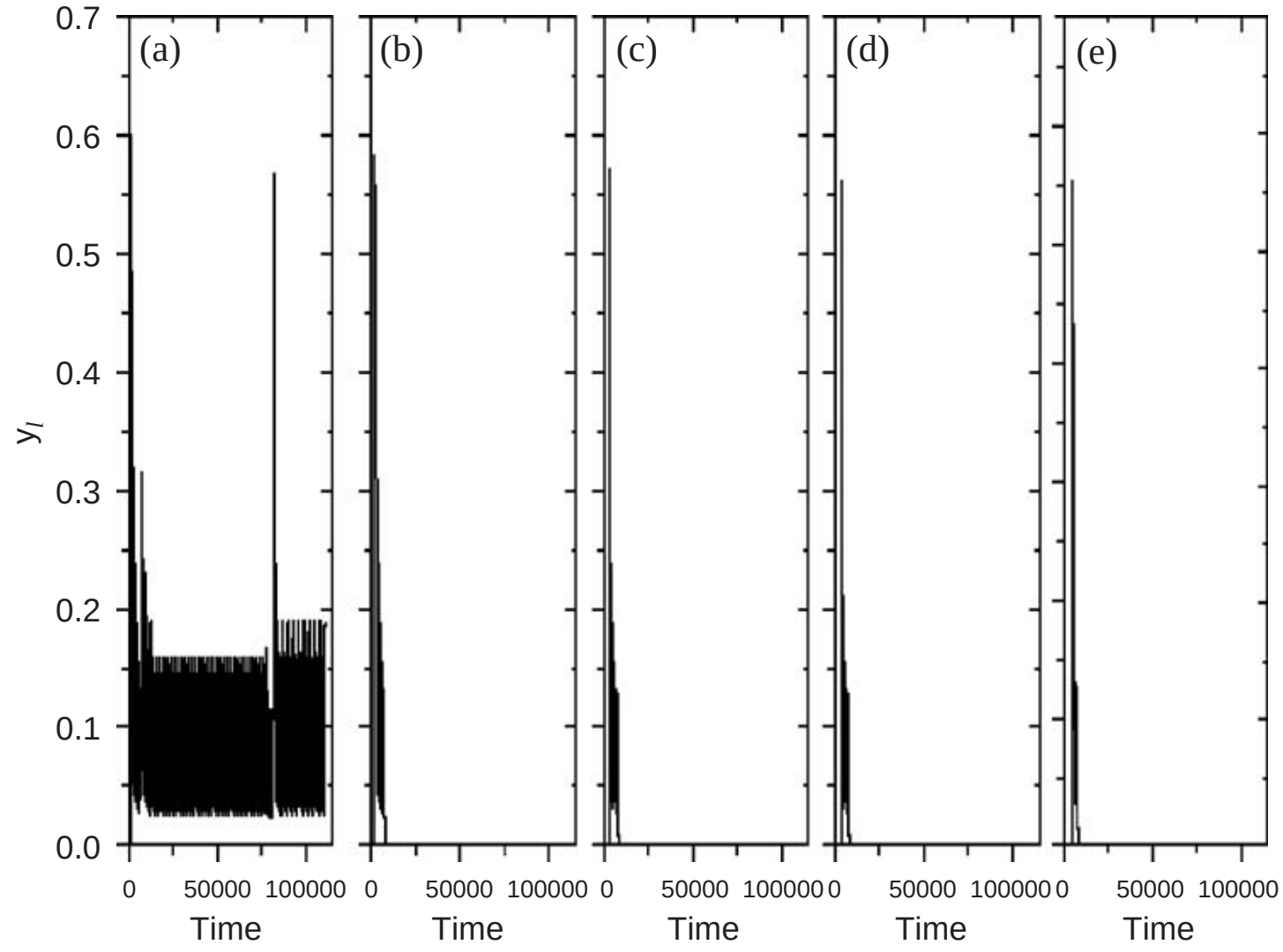

Fig. 11. Performance of the model for the first five populations of clones that recognize a specific population of antigen (concentration equal to 0.10), shown individually (a–e). All the populations (1–5) are excited after the inoculation, however just the first population of clones (a) remains for a long time after the suppression of the antigen.

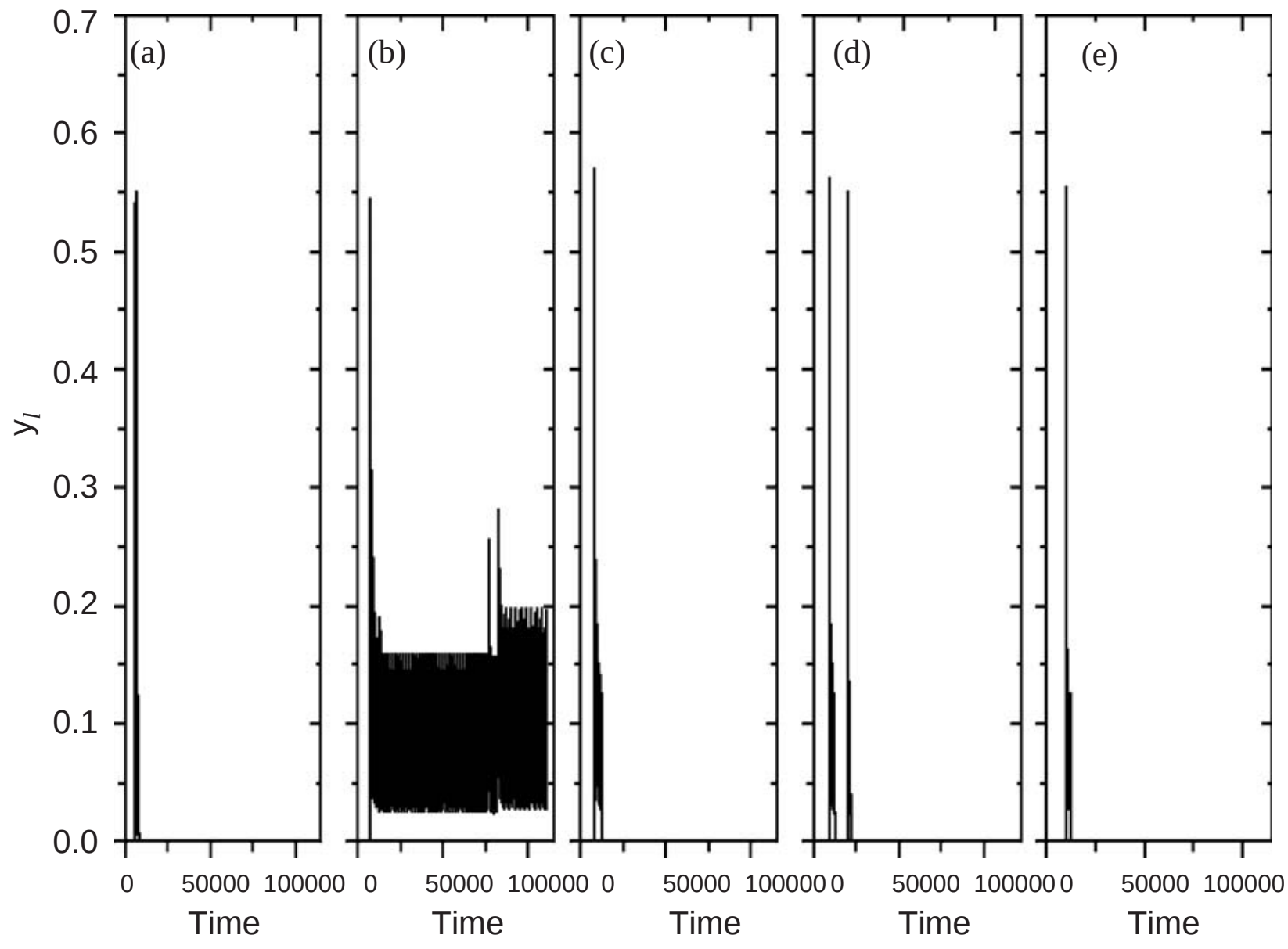

Fig. 12. Performance of the model for the first five populations of clones that recognize a specific population of antigen (concentration equal to 0.10), shown individually (a–e). All the populations (6–10) are excited after the inoculation, however just the seventh population of clones (b) remains for a long time after the suppression of the antigen.

Figs. 15(a) and (b) show that the global behavior of the lifetime for the populations of clones that recognize the antigens varies when the antigen concentration is changed. Therefore it is impossible to predict which population will survive for a long period. These results are in agreement with the fact that it is impossible to foresee the validity of a vaccine in a healthy individual.

## 4. Discussion and conclusions

In this paper, an extension of the model by Lagreca et al. [15] is proposed, to include antibodies response, in the study of the dynamics of an infectious disease, considering structural regulation mechanisms. In the approach presented here, we define clones as an ensemble of B-cells, and the populations of antibodies are treated separately. Consequently we have obtained that the time evolution of the clones is different from the time evolution of the populations of secreted antibodies, in agreement with what is expected for a normal immune system. The fundamental role of the antibodies in the mediation of the overall control of the differentiation of the B-cells was considered, showing that they affect globally the immunological memory. This model considers that, in dynamic equilibrium, the cell–cell interactions [21,22] result in memory maintenance. Another important feature is that the duration of the immunological memory is not the same for all the injections, some populations of clones having longer life than others.

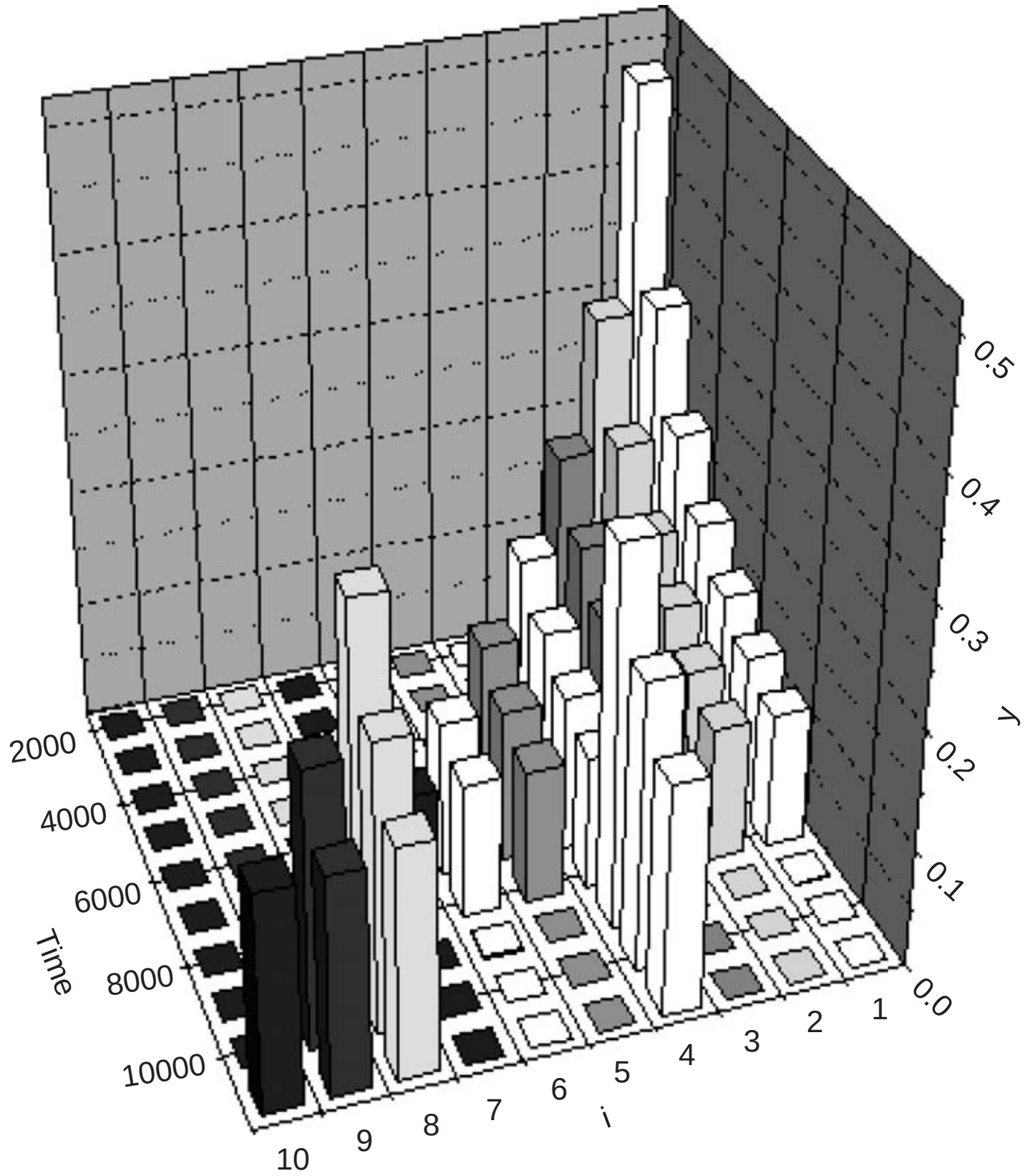

Fig. 13. Evolution of the populations of clones that survive 11 000 time steps, for the concentration of antigens equal to 0.08. The fourth and the eighth populations remain excited. The other populations of clones die except for the last population which were excited close to the end of period.

Although the duration of the memory for specific antigen to be random, our results suggests that the decrease of the production of antibodies favors the global maintenance of immune memory; high production of antibodies is useful to combat infections, however it is harmful for the memory clonal populations produced by previous infections.

In this model there is no need of persistent antigen or the existence of long-living memory lymphocytes [23–25]. The presence of Burnet cells and of complementary Jerne cells originate the memory regenerating system through idiotypic–antiidiotypic interactions of their surface immunoglobulins, what leads to a self-perpetuation [16,21,22].

We have considered the immune system as a network of molecules and cells that recognize one another even in the absence of antigen. Since the antibodies are generated by random genetic mechanisms, they can be regarded as strangers by the rest of the body and be treated as antigens. The same occurs for the clone receptors that can react among them and with the antibodies. The cells that recognize the antigens and their clonal derivatives (Burnet cells) select a pool of naive complementary clones (Jerne cells) that can react with the idiotypes of Burnet cells. Thus, when Burnet cells and Jerne cells interact, clonal expansion of these complementary cells take place. In our simulation, this happens when an antigen is inoculated in the system and two populations are excited: the population of cells that recognize the antigen and its image. The activation of clones, through the idiotypic–antiidiotypic interaction, results in the proliferation and recruitment of higher affinity cells. In conclusion, clonal memory can be considered as a function propagated by the interaction of a series of Burnet cells and Jerne cells.

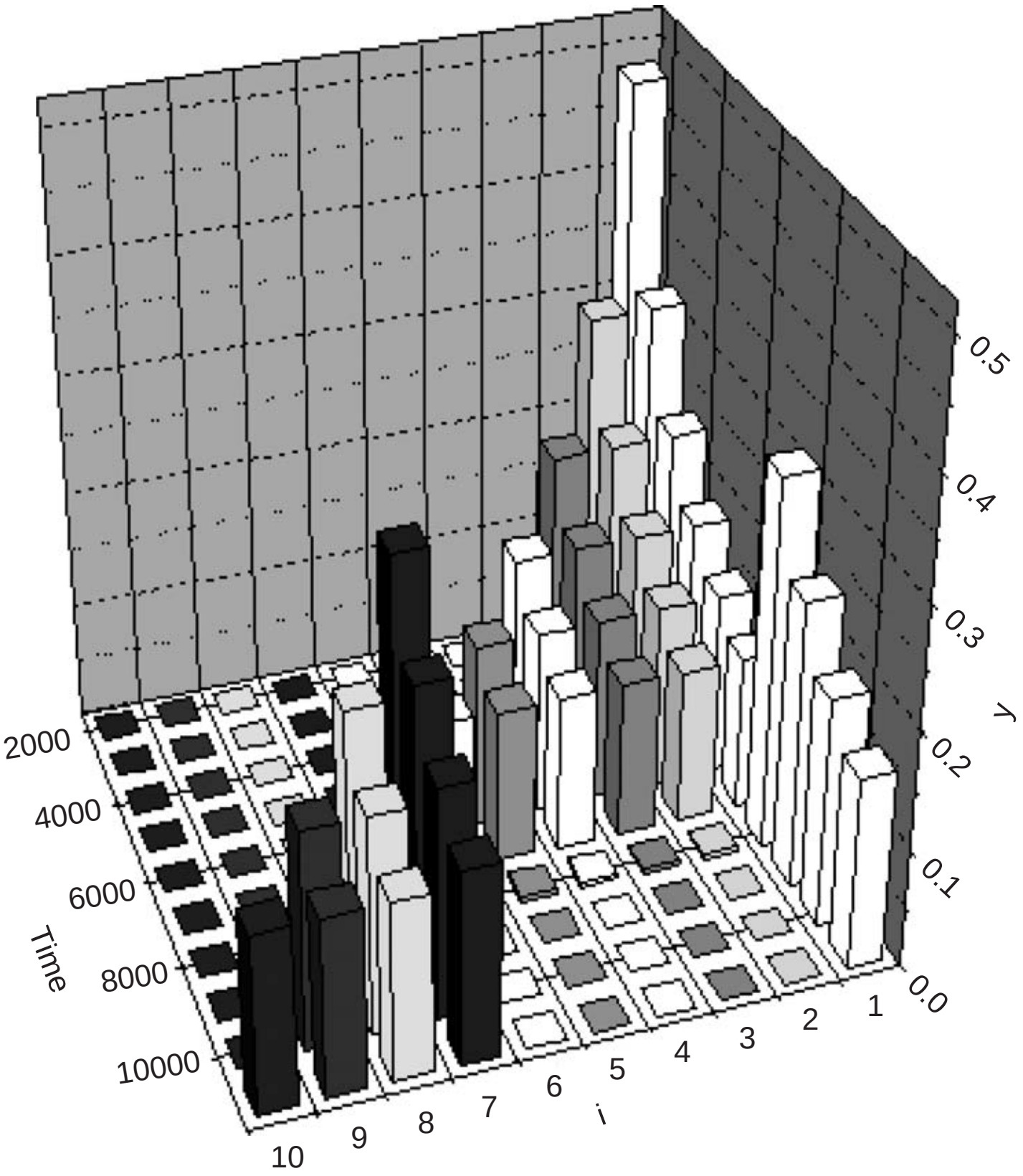

Fig. 14. Evolution of the populations of clones that survive 11 000 time steps, for the concentration of antigens equal to 0.10. The first and the seventh populations remain excited. The other populations of clones die except for the last population which were excited close to the end of period.

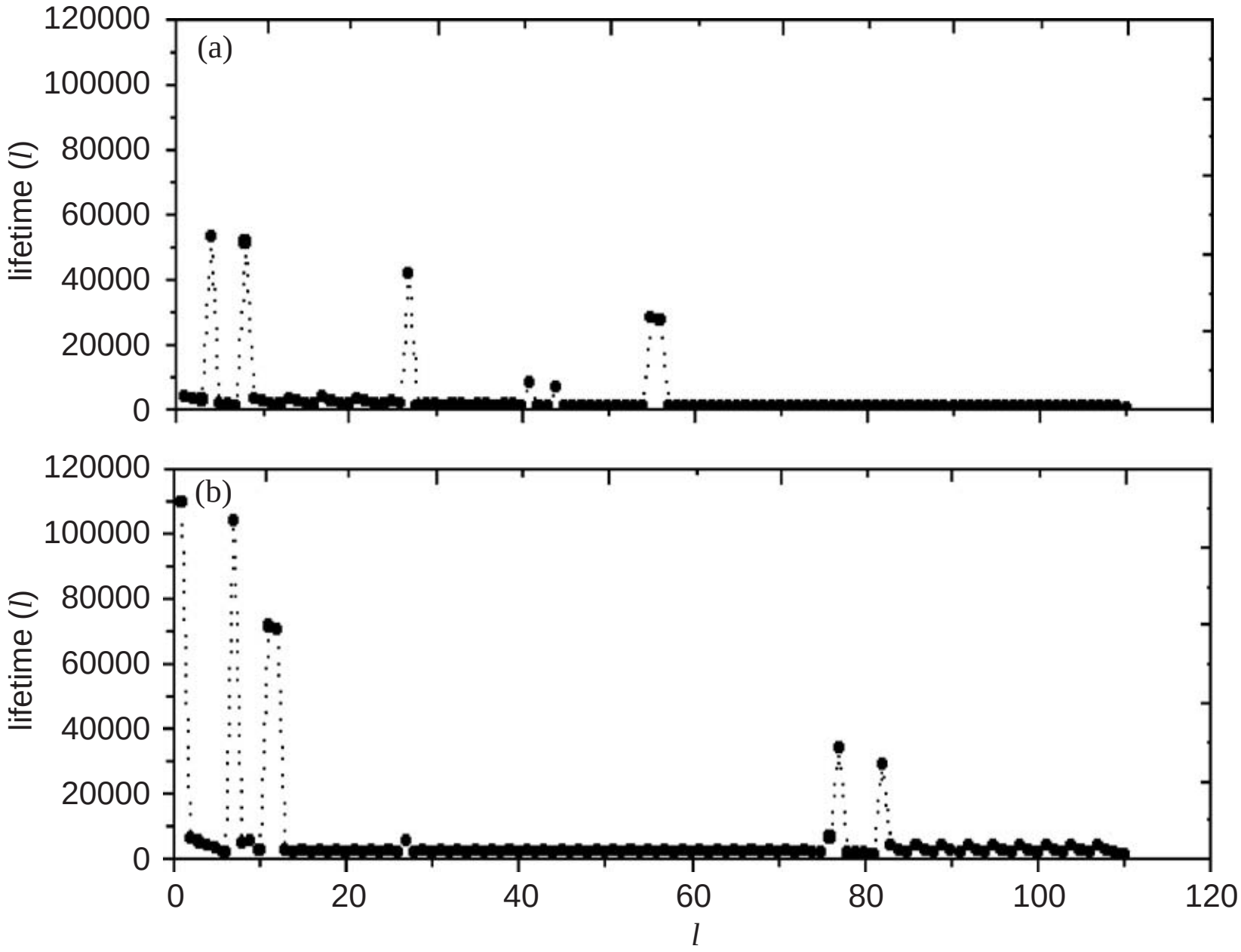

Fig. 15. Lifetime for the populations of clones that recognize each specific antigen. The concentration of antigens is equal to 0.08 in (a) and 0.10 in (b).

## Acknowledgements

This work was partially supported by brazilian agency CNPq. The author wishes to thank Dr. V.B. Campos for valuable discussions.

## References

[1] I. Roitt, J. Brostoff, D. Male, Immunology, fifth ed., Mosby, New York, 1998.

[2] A.K. Abbas, A.H. Lichtman, J.S. Pober, Cellular and Molecular Immunology, second ed., W.B. Saunders Co, London, 2000.

[3] B. Alberts, D. Bray, J. Lewis, M. Raff, K. Roberts, J.D. Watson, Molecular Biology of the Cell, fourth ed., Garland Publishing, New York, 1997.

[4] J.H.L. Playfair, Immunology at a Glance, sixth ed., Blackwell Scientific Publications, Oxford, 1996.

[5] O. Levy, Eur. J. Haematol. 56 (1996) 263.

[6] M. Reth, Immunol. Today 16 (1995) 310.

[7] G. Moller, Immunol. Rev. 153 (1996).

[8] F.M. Burnet, The Clonal Selection Theory of Acquired Immunity, Vanderbuilt University, Nashville, TN, 1959.

[9] N.K. Jerne, Clonal selection in a lymphocyte network, in: G.M. Edelman (Ed.), Cellular Selection and Regulation in the Immune Response, Raven Press, New York, 1974, p. 39.

[10] N.K. Jerne, Ann. Immunol. 125 (1974) 373.

[11] F. Celada, P.E. Seiden, Immunol. Today 13 (1992) 53.

[12] F. Celada, P.E. Seiden, J. Theoret. Biol. 158 (1992) 329.

[13] D. Mopurgo, F. Celada, P.E. Seiden, Int. Immunol. 7 (1995) 4.
[14] D. Mopurgo, F. Celada, P.E. Seiden, Eur. J. Immunol. 7 (1996) 1350.
[15] M.C. Lagreca, R.M.C. de Almeida, R.M. Zorzenon dos Santos, Physica A 289 (2000) 42.
[16] R. Nayak, S. Mitra-Kaushik, M.S. Shaila, Immunology 102 (2001) 387.
[17] M. Kikuchi, Y. Okabe, Phys. Rev. B 35 (1995) 5382.
[18] M. Kikuchi, Y. Okabe, Int. J. Mod. Phys. C 7 (1995) 747.
[19] G. Bhanot, D. Duke, R. Salvador R, Phys. Rev. B 33 (1986) 7841.
[20] C. Michael, Phys. Rev. B 33 (1986) 7861.
[21] A.S. Perelson, G. Weisbush, Rev. Mod. Phys. 69 (1997) 1219.
[22] A.S. Perelson, R. Hightower, S. Forrest, Res. Immunol. 147 (1996) 202.
[23] D. Gray, M. Kosko, B. Stockinger, Int. Immunol. 3 (1991) 141.
[24] D. Gray, Ann. Rev. Immunol. 11 (1993) 49.
[25] C.R. Mackay, Adv. Immunol. 53 (1993) 217.